\documentclass[11pt,twoside]{article}
\usepackage{./asp2010}

\resetcounters

\newcommand{\msune}{M$_{\odot}$} 
\newcommand{\hi}{H{\footnotesize I} }
\newcommand{\hie}{H{\footnotesize I}}
\newcommand{\dgr}{$^{\circ}~$}

\begin{document}

\title{Disk Destruction and (Re)-Creation in the Magellanic Clouds}
\author{David L. Nidever,$^1$}
\affil{$^1$Department of Astronomy, University of Michigan, 1022 Dennison Bldg., 500 Church St., Ann Arbor, MI 48109, USA}

\begin{abstract}
Unlike most satellite galaxies in the Local Group that have long lost
their gaseous disks, the Magellanic Clouds are gas-rich dwarf galaxies
most-likely on their first pericentric passage allowing us to study disk
evolution on the smallest scales.  The Magellanic Clouds show both
disk destruction and (re)-creation.  The Large Magellanic Cloud has a
very extended stellar disk reaching to at least 15 kpc (10 radial
scalelengths) while its gaseous disk is truncated at $\sim$5 kpc mainly due
to its interaction with the hot gaseous halo of the Milky Way.  The
stellar disk of the Small Magellanic Cloud, on the other hand, has
essentially been destroyed.  The old stellar populations show no sign
of rotation (being pressure supported) and have an irregular and
elongated shape.  The SMC has been severely disturbed by its close
encounters with the LMC (the most recent only 200 Myr ago) which have
also stripped out large quantities of gas creating much of the
Magellanic Stream and the Magellanic Bridge.  Amazingly, the SMC has
an intact, rotating \hi disk indicating that either the inner \hi
was preserved from destruction, or, more likely, that the \hi disk
reformed quickly after the last close encounter with the LMC.
\end{abstract}

{\bf Background:}
Essentially no Local Group (LG) dwarfs within 270 kpc of their host galaxy (Milky Way or M31) currently have \hi
\citep[$<$10$^5$ \msune;][]{Grcevich09}.
The lone exceptions are the Magellanic Clouds (MCs).  It is thought that the gas was ram pressure stripped from the
inner galaxies, but it is complicated to study now.  Fortunately, the MCs can be used to study these gas-dynamical
effects right now in great detail because the MCs still have a lot of gas \citep[likely falling into the Milky Way
for the first time;][]{Besla07} and are quite nearby.

One of the most striking features of the Magellanic system is the 200\deg--long Magellanic Stream \citep[MS; ][]{Nidever10}.
The MCs have lost lots of gas \citep[$\sim5\times10^8$ \msune;][]{Bruens05} to the MS and Leading Arm in
the last couple of Gyrs due to their interaction with the MW and each other.  The MS is spatially bifurcated into
two filaments \citep{Putman03} and one of the filaments can be traced back to its origin in the LMC using its
velocity coherence \citep{Nidever08}.  $HST$ absorption line metallicities show that the MS has a dual origin with
one filament coming from the LMC and the other from the SMC \citep{Fox13,Richter13}.  Recent models indicate
that the SMC gas was tidally stripped by the LMC in its most recent incounters \citep{Besla10,DB12}.

\begin{figure}[t]
$\begin{array}{cc}
\fbox{\includegraphics[trim=20mm 18mm 22mm 0mm, clip=true, scale=0.30]{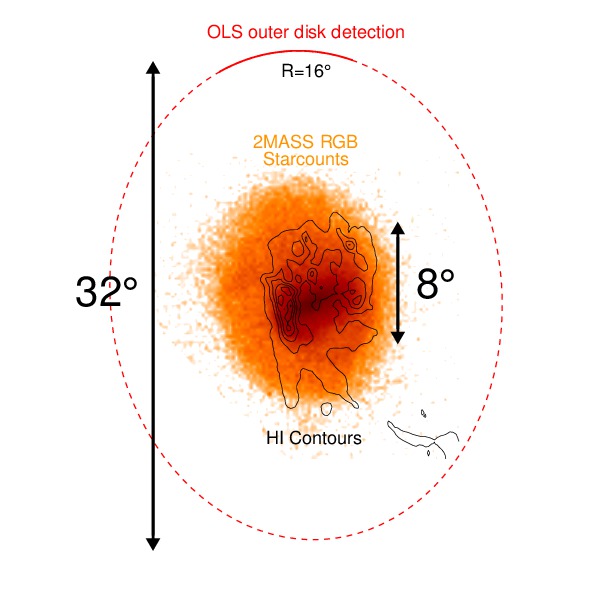}} &
\includegraphics[angle=0.0,scale=0.28]{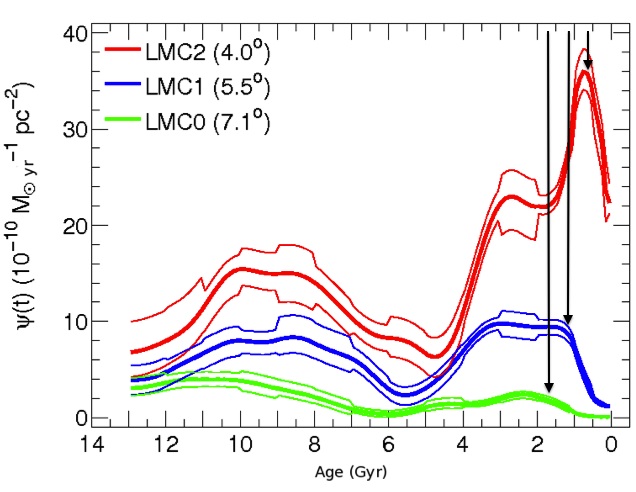}
\end{array}$
\caption{(Left) Map of the LMC showing the 2MASS RGB starcounts in orange and the \hi contours in black.  The
red ellipse shows the extent of the LMC stellar disk seen by the Outer Limits Survey (OLS, Saha et al. 2010).
The stellar disk is substantially larger (32\deg) than the \hi disk (8\deg).
(Right) The star formation history of three LMC fields (Gallart et al.\ 2009; Meschin et al. 2013).
In all three fields the star formation rate has dropped dramatically in the recent past (indicated
by vertical arrows).  The drop started at the most distant field first and then moved inward over time.
The innermost field is at the outer edge of the LMC \hi disk.  More central fields show ongoing star
formation.}
\label{fig_lmc}
\end{figure}

{\bf The Large Magellanic Cloud:}
The Large Magellanic Cloud (LMC) has an inclined, exponential, rotating stellar disk \citep{vdM01,vdM02,vdM13}.
It also has an off-center stellar bar \citep{vdM01}, a high star formation rate \citep{SH02}, and a rotating \hi
disk \citep{Kim98}.  The LMC stellar and gaseous distributions are quite different.  The stellar disk is regular,
exponential and slightly elliptical, while the gaseous disk is smaller, rectangular, has a (morphological
center that is offset from the stellar center, and also has a high column density on the leading edge (see Fig.\
\ref{fig_lmc} left panel).  The Outer Limits Survey (OLS) used deep photometry of old MSTO stars to trace the LMC disk
to $R$=16\dgr and showed that it still followed an exponential out to these large radii \citep{Saha10}.
Spectroscopically-detected LMC giants have been traced to $R$$\approx$20\dgr over 180\dgr in position angle
\citep{Munoz06,Majewski09} and in some regions out to $R$$\sim$27\dgr \citep{Munoz13}.  However, the density
profile and radial velocities of these stars are more consistent with a halo than a disk.
Nevertheless, the current \hi disk (8\deg) is puny compared to the stellar disk (32\deg) and it appears as if
the gaseous disk has been whittled away.  But by what?  Whatever the cause it can't have affected the stars
that much because the LMC hosts an intact extended stellar disk.

The LMC star formation histories provide further evidence of the gaseous disk destruction.  The right panel of 
Figure \ref{fig_lmc} shows the star formation histories for three intermediate radius LMC fields
\citep[$R$=4.0\deg, 5.5\deg, and 7.1\deg;][]{Gallart09,Meschin13}.  Stars were forming in all three fields within
the last few Gyrs.  However, recently there has been a severe drop in the star formation rate (SFR) starting
with the outermost field and then moving inward over time.  The innermost field (at the edge of the \hi disk in
a supergiant shell) has current ongoing star formation but if the decline in SFR continues this star formation will
cease with less than one Gyr.  The outer two fields where star formation has essentially ceased are outside the
current \hi disk.  The rate at which the cutoff is moving inward is $\sim$2.8\deg/Gyr=2.4 kpc/Gyr.  At this rate,
star formation in the LMC will completely cease within $\sim$0.8 Gyr and, presumably, also the \hi disk will be gone.

The six mechanisms could that be destroying the LMC gaseous disk are: 1) star formation (gas$\rightarrow$stars),
2) stellar feedback (gas), 3) MW tidal stripping (gas/stars), 4) MW ram pressure stripping (gas), 5) SMC
tidal stripping (gas/stars), and 6) SMC ram pressure stripping (gas).  The LMC has had continuous star formation
for $\sim$12 Gyr even though the gas depletion rates for dwarfs are normally $\sim$2 Gyr \citep{Bigiel08}.  Gas
accretion is required to keep the star formation going for this long, and it is quite unlikely that this would
all stop right now (rules out \#1).  As previously mentioned, the stellar disk is basically unaffected, so the mechanism
needs to be something that only affects gas.  This rules out tidal forces (\#3 and \#5).
How about the LMC/SMC interaction?  The LMC and SMC have had a recent ($\sim$200) close encounter \citep{Ruzicka10}
and \citet{Besla12} even suggest that there was a direct collision that can explain the LMC off-center bar.  could
this explain the outer LMC \hi disk destruction?  The collsion was too recent (100--300 Myr ago) to explaine the
outer disk destruction ($\sim$2 Gyr ago to present) and it also can't explain the LMC gas in the MS downstream.  So
So although a direct collision might have occurred, it isn't the whole story. There must be something else affecting
the LMC HI disk (rules out \#6).  Finally, we are left with two mechanisms: MW ram pressure stripping, and stellar
feedback.  Ram pressure is consistent with the boxy shape of the \hi disk, large column density gradient on the leadging
edge, the timescale of LMC recent falling into the MW, as well as the \citet{Mastropietro05} result showing you can
get significant ram pressure stripping from the LMC.  Furthermore, stellar feedback and outflow is consistent with
the large number of supergiant shells in the LMC \citep{Kim98}, gaseous outflow in the northeast of the LMC
\citep{Kim98,Nidever08}, and the shutoff of star star formation after a ``burst'' as seen in the 4.0\dgr field.

{\bf The Small Magellanic Cloud:}
\citet{Gardiner92} used photographic plate photometry to study the SMC periphery and showed that
the disribution of young stars is very irregular and elongated towards the LMC.  This elongated
distribution is also seen in \hie, the Magellanic Bridge \citep{Muller03}, that was recently
tidally stripped from the SMC \citep{MB07}.  The distribution of the red (and older) stars is much
more regular and symmetric extending to $R$$\sim$4--5\dgr \citep{Gardiner92}.  \citet{Nidever11}
used data from the MAgellanic Periphery Survey (MAPS) to trace photometrically-selected SMC giant
stars to $R$$\sim$11\dgr and showed that they follow a slightly elliptical exponential
profile out to $R$$\sim$8 kpc.  It is well known that the SMC has a large line-of-sight depth \citep{HH89}.
\citet{Nidever13} derived distances for red clump stars in eight fields at $R$=8\dgr and found a
bimodality in distance in the eastern fields with a new structure at $\sim$55 kpc on the near-side
of the SMC.  The most likely interpretation of this new structure is that it is the stellar counterpart
of the tidally stripped Magellanic Bridge.

There is no detected rotation in the SMC stellar distribution either in radial velocities \citep{Kunkel00,HZ06}
or proper motions \citep{Piatek08}, instead the stars appear to be pressure supported.  Furthermore, the line-of-sight
depths of red clump stars with radius on the western side of the SMC are more consistent with a spheroidal
distribution than a disk distribution \citep{GH91}.  The evidence, therefore, points to the SMC having
a spheroidal-like (or ellipsoidal) structure rather than an exponential disk.  If there was a SMC stellar disk
before (which is likely) then it was severally disturbed by the recent encounter with the LMC.
On the other hand, rotation {\em is seen} in the SMC \hi distribution \citep{Stani04}.
How is it possible for the stars to be so perturbed (with no rotation) but the gas to be rotating?
Recollapse into a disk.  The stars are collisionless and non-dissipational and once the stars are stirred up
and perturbed they can't easily relax to a disk again.  The gas, on the other hand, is collisional and
dissipational and can recollapse to a rotating disk even after being disturbed.  The SMC free-fall timescale is 
$\sim$50 Myr. Since the LMC/SMC collision happened 100--300 Myr ago that should give the gas enough time to collapse.

In the near future, the Survey of the MAgellanic Stellar History (SMASH, PI:Nidever), an approved community DECam project
to perform deep imaging of the MCs and their periphery, will provide exquisite spatially-resolved star formation
histories that will shed new light on the complex history and evolution of the Magellanic Clouds.
 
\acknowledgements
I would like to thank Carme Gallart, Sne\u{z}ana Stanimirovi{\'c}, Jay Gallaugher, Matt Haffner, Gurtina Besla,
Knut Olsen, and Noelia Noel for useful conversations about the structure and evolution of the Magellanic Clouds.

\end{document}